\documentclass[a4paper,UKenglish,cleveref, autoref, thm-restate]{lipics-v2021}

\hideLIPIcs
\nolinenumbers
\usepackage{graphicx} 
\usepackage{amsmath}
\usepackage{amssymb}

\usepackage{algorithm}
\usepackage{algpseudocode}
\usepackage{xspace}
\usepackage{caption}

\newcommand{\fon}{f^{\mbox{\footnotesize \it on}}}
\newcommand{\fdc}{f^{\mbox{\footnotesize  \it dc}}}
\newcommand{\foff}{f^{\mbox{\footnotesize  \it off}}}

\newcommand{\zdue}{\ensuremath{\mathbb{Z}/(2)}\xspace}
\newcommand{\zp}{\ensuremath{\mathbb{Z}/(p)}\xspace}
\newcommand{\fc}{\ensuremath{\mathbb{F}_c}\xspace}

\newcommand{\bigspace}{\ensuremath{\cal{V}}\xspace}
\newcommand{\solution}{{\ensuremath{\cal{W}}\xspace}}
\newcommand{\inputset}{\ensuremath{\cal{S}}\xspace}

\newcommand{\defproblem}[3]{
\vspace{2mm}
\noindent\fbox{
   \begin{minipage}{0.96\textwidth}
   \textsc{#1}\\
   {\bf{Input:}} #2  \\
   {\bf{Output:}} #3
   \end{minipage}
   }
   \vspace{2mm}
}

\title{Efficient Enumeration of Enclosed Vector Spaces} 

\author{Anna Bernasconi, Alessio Conte, Giulia Punzi}{University of Pisa, Italy}{}{}{}

\author{Valentina Ciriani}{Università degli Studi di Milano, Italy}{}{}{}

\author{Alberto L'Episcopo}{Scuola Normale Superiore, Pisa, Italy}{}{}{}

\authorrunning{Bernasconi et al.} 

\Copyright{Bernasconi et al.} 



\category{} 

\relatedversion{} 

\EventEditors{John Q. Open and Joan R. Access}
\EventNoEds{2}
\EventLongTitle{42nd Conference on Very Important Topics (CVIT 2016)}
\EventShortTitle{CVIT 2016}
\EventAcronym{CVIT}
\EventYear{2016}
\EventDate{December 24--27, 2016}
\EventLocation{Little Whinging, United Kingdom}
\EventLogo{}
\SeriesVolume{42}
\ArticleNo{23}

\begin{document}

\maketitle

\begin{abstract}
In this paper, we address several problems concerning vector spaces enclosed in a given set. Let $\bigspace$ be a vector space over a finite field of cardinality $c$, and let $\inputset \subseteq \bigspace$ be a set of vectors. A space enclosed in $\inputset$ is a vector subspace $\solution$ of $\bigspace$ that is also contained in $\inputset$: $\solution \subseteq \inputset$.
We focus on enumeration problems, where the task is to list all solutions, and we first provide an algorithm to enumerate all spaces that are enclosed in $\inputset$.
Our algorithm is further adapted to solve two more problems: the enumeration of (inclusion-)\emph{maximal} enclosed spaces, and the problem of finding an enclosed space of \emph{maximum} dimension. The latter problem arises in the context of Boolean functions' regularity detection. It can also be seen as a dual version of the well-known linear span: indeed, the span is the minimum-dimension vector space that contains a given set of vectors $\inputset$, and it is a fundamental concept in linear algebra. 
Our proposed algorithms are based on the binary partition paradigm, and have total time complexity 
$e^{\frac{1}{2\ln c}\ln^2 n - \Theta(\log n \log \log n)}$, where $n= |\inputset|$. The first version, for enumerating all enclosed spaces, also achieves a delay (time between consecutive outputs) of $O(n)$. Our algorithms provide a quadratic speed-up with respect to a brute-force approach, although the speed-up appears even greater in our experimental evaluation on boolean vector spaces.
\end{abstract}

\section{Introduction}
\label{intro}
In mathematical and computational contexts, certain problems require, or allow,  the domain augmentation, i.e., the introduction of additional points to the input data, to enhance structural regularity among them, and thereby improve the efficiency of their algorithmic processing. For instance, the computation of the linear span of a given set of vectors is a typical case in which a set is enriched with other elements so that it assumes the algebraic structure of a vector space. 
Given a vector space ${\bigspace}$ and a subset ${\inputset} \subseteq {\bigspace}$, the computation of the linear span of ${\inputset}$---the smallest (and unique) vector space ${\solution} $ s.t. ${\inputset} \subseteq {\solution} \subseteq {\bigspace}$---simply consists in selecting the maximum number of linearly independent vectors in ${\inputset}$ that represent the basis of ${\solution}$. 
This problem can be solved efficiently in polynomial time applying the well-known Gauss-Jordan elimination procedure~\cite{L03}. 

In contrast, certain problems might benefit from a domain reduction, where the selective removal of elements can enhance the structural regularity of the domain.  The problems addressed in this study— concerning spaces enclosed in a given set of vectors $\inputset$ ---fall  precisely into this category. Indeed, to solve these problems, it is necessary to select elements from ${\inputset}$ whose removal transforms ${\inputset}$ into a vector space.

Specifically, in this paper we consider the problem of enumerating all vector spaces enclosed in $\inputset$, together with two additional problems: the enumeration of \emph{maximal} enclosed spaces, with respect to inclusion, and the problem of finding a space of \emph{maximum} dimension contained in $\inputset$\footnote{Observe that such space is not necessarily unique, see Example~\ref{es1}.}. This latter problem, which arises, for example, in the context of Boolean functions' regularity detection, can also be viewed as a dual version of the computation of the linear span of ${\inputset}$: while the span operation expands a set to its smallest containing space, our focus here is on reducing a set to a largest enclosed space. 

To the best of our knowledge, this problem, while quite natural, has not been explicitly addressed in the existing literature, where one can instead find references to the somewhat related problem of counting all subspaces of a given vector space over a finite field, a problem that has been discussed and solved long ago. Indeed, it is well known that the number of $k$-dimensional vector subspaces of an $n$-dimensional vector space over 
a finite field of cardinality $q$ is given by the {\em Gaussian binomial coefficient} 
$\binom{n}{k}_q$
(see~\cite{GR69,K71} for more details). 

It is important to note that the problems addressed in this paper differ from this classical one in mainly two ways: first, we start from a set rather than a vector space; moreover, we are interested not only in counting all enclosed spaces, but in enumerating\footnote{In this context, enumeration refers to the problem of listing all of the solutions, rather than just counting their number} the maximal ones and in finding an enclosed space of maximum dimension. Note that the latter two problems obviously become trivial when the starting set ${\inputset}$ is already a vector space, instead of just a set of vectors.

Our main contribution consists of the design and experimental evaluation of an enumeration algorithm based on the binary partition paradigm. 
The proposed algorithm has quasi-polynomial time complexity, theoretically providing a quadratic speed-up compared to a brute-force approach. This speed-up has proven to be even more significant in the experimental evaluation of the algorithm.

The proposed algorithm, particularly in its version related to identifying the largest enclosed space, can be exploited to derive structural regularities in Boolean functions~\cite{CMCD25}. 
An explicit example of application related to the discovery of structural regularities
is presented and discussed in Appendix~\ref{section:motivation}. This application arises from the idea of extending a structural property known as autosymmetry from completely specified Boolean functions to incompletely specified ones. As discussed in Appendix~\ref{section:motivation}, to maximize this structural regularity of a target function $f$, depending on $\ell$ binary variables, it is necessary to ``extract'' a vector space $L_f$ of maximum dimension from the so-called {\em closure set} $S_f$ of $f$. 
Thus, this task involves the selective removal of elements from a given set of vectors to enhance the regularity of the domain, and can be solved applying the algorithm for identifying the largest enclosed space, with 
${\bigspace} = \{0,1\}^\ell$ (viewed as a vector space over \zdue), ${\inputset} = S_f \subseteq \cal V$,  and ${\solution}= L_f$.

Another interesting application involves deriving approximated regular Boolean functions~\cite{DWLSXNLQ25,SACPR20} in the context of approximate logic synthesis. This trend in logic synthesis allows modifications to certain outputs of a logic specification, within the error tolerance of a given application, to reduce the complexity of the final circuit implementation. Specifically, the proposed algorithm could be applied in scenarios where only ``under-approximation'' of functions is allowed, meaning that the outputs can only be changed from 1 to 0.

\section{Problem Definition and Paper Contribution}
This paper deals with vector spaces and subspaces over fields, we give a brief overview of their basic definitions in Appendix~\ref{subsection:fields-vecspace}. 

Let us consider a vector space \bigspace over a finite field \fc of cardinality $c$.
Let $\inputset$ be a subset of $\bigspace$ of cardinality $n$. We say that a vector subspace $\solution$ of $\bigspace$ is \emph{enclosed} in $\inputset$ if $\solution\subseteq \inputset$ (as sets). 
In an enumeration problem, we are asked to output all the solutions to a given problem. In this work, we are interested in the {enumeration} of several kinds of spaces enclosed in the set \inputset.
The first problem studied in this paper is that of enumerating all subspaces $\solution$ of $\bigspace$ contained (or enclosed) in $\inputset$, formally stated as follows: 

\defproblem{\textsc{Enclosed Space Enumeration}}{A set of vectors ${\inputset} \subseteq {\bigspace}$}{All vector spaces ${\solution} \subseteq {\inputset}$ }

When considering all the vector spaces contained in $\inputset$, we are handling a lot of redundancy: let $\solution$ be a vector space and $\mathcal{B}_{\solution}= \{v_1,...,v_k\}$ a basis, then any distinct subset of vectors of $\mathcal{B}_{\solution}$ will be the basis of a different vector space $\solution ' \subsetneq \solution$. In this situation, we may want to discard $\solution '$ to reduce the output size without losing relevant information. Thus, we also consider the problem of enumerating the inclusion-maximal enclosed spaces: an enclosed space $\solution$ is {\em maximal} if there does not exist any vector space $\mathcal{Z} \subseteq {\inputset}$ such that $\solution\subsetneq \mathcal{Z}$.

\defproblem{\textsc{Maximal Enclosed Space Enumeration}}{A set of vectors ${\inputset} \subseteq {\bigspace}$}{All maximal vector spaces ${\solution} \subseteq {\inputset}$ }

We note that not all of the maximal spaces enclosed in a set ${\inputset}$ have the same dimension (see Example~\ref{es1}). Thus, we finally consider the following problem of separate interest:

\defproblem{\textsc{Maximum Enclosed Space}}{A set of vectors ${\inputset} \subseteq {\bigspace}$}{A vector space ${\solution} \subseteq {\inputset}$ of maximum dimension }

As already observed in Section~\ref{intro}, the \textsc{Maximum Enclosed Space} problem can be viewed as 
a dual version of the computation of the span of a set of vectors. However, the two problems also differ deeply. 
First of all, the span of a set ${\inputset}$ of vectors is unique, while, as shown in the example below, ${\inputset}$ can contain different spaces of maximum dimension.
Furthermore, while we have polynomial-time algorithms available for computing the span, to the best of our knowledge, our proposed algorithm is the first one to achieve quasi-polynomial time complexity for the \textsc{Maximum Enclosed Space} problem.

\begin{example}\label{es1}

    Let ${\bigspace} = \{0,1\}^4$, viewed as a vector space of dimension 4 over \zdue, and consider the subset ${\inputset} = \{(0,0,0,0),(0,0,0,1),(0,0,1,1),(0,0,1,0),(0,1,0,0),(0,1,1,0),(1,0,0,0)\}$, of cardinality 7. Observe that ${\inputset}$ contains the zero vector, but it is not a vector space. The three problems would have output as follows:
    \begin{itemize}
        \item \textsc{Enclosed Space Enumeration} would output the 8 spaces contained in $\inputset$: 6 one-dimensional spaces, each represented by pairs consisting of the zero vector and a non-zero vector, and 
         2 two-dimensional spaces  ${\solution}_1 = \{(0,0,0,0),(0,0,0,1),(0,0,1,0),(0,0,1,1)\},$  ${\solution}_2 = \{(0,0,0,0),(0,0,1,0),(0,1,0,0),(0,1,1,0)\}\,;$
         
        \item Of the previous spaces, \textsc{Maximal Enclosed Space Enumeration} outputs the following 3 maximal spaces:  ${\solution}_1$,  ${\solution}_2$, and the one-dimensional space  ${\solution}_3 = \{(0,0,0,0),(1,0,0,0)\}$;        
        
        \item Finally, \textsc{Maximum Enclosed Space} would output one of the 2 spaces of maximum dimension:  ${\solution}_1$ and ${\solution}_2$.
     \end{itemize}
     Finally, observe that in this case $\text{Span}({\inputset}) = {\bigspace}$.
\end{example}

\smallskip

To the best of our knowledge, the proposed problems, albeit natural, have not been previously studied in the literature.
In the present paper, we start by describing a brute-force approach to solve the proposed problems, and studying its complexity. We then present a faster algorithm for the \textsc{Enclosed Space Enumeration} problem, which can be adapted to solve the other two problems with the same complexity. More formally, we prove the following:
\begin{theorem}
   The \textsc{Enclosed Space Enumeration}, \textsc{Maximal Enclosed Space Enumeration}, and \textsc{Maximum Enclosed Space} problems for an input set $\inputset$ of size $n$ can be solved in 
   $e^{\frac{1}{2\ln c}\ln^2 n - \Theta(\log n \log \log n)}$
   total time using $O(n)$ space. 
\end{theorem}

In enumeration problems, the size of the output is often exponential in the input size, and thus total time complexity is usually exponential. For this reason, in this field there are additional notions of \emph{output-sensitive} complexities, where the complexity of an algorithm is sometimes given with respect to the output size, instead of the input size.
One such complexity notion is the \emph{delay}: the delay of an enumeration algorithm is the time that passes between an output and the next one. We also provide a result about the delay of the first version of our algorithm:

\begin{proposition}
\textsc{Enclosed Space Enumeration} can be solved in $O(n)$ delay. 
\end{proposition}

In the rest of the paper, we study the problems for the boolean vector space \bigspace, that is, the vector space over the field \zdue of integers modulo 2. 
We later show how our proposed solutions can be easily generalized from the Boolean case to vector spaces over any finite field \fc 
(more on this in Section~\ref{section:bp-generalization}). So, from now on, unless otherwise stated, we consider the set \bigspace$= \{0,1\}^\ell$ (binary $\ell$-tuples) equipped with the structure of a vector space over the field $\zdue$ (i.e., vector addition is the boolean XOR).

We will further assume that $\ell$ is constant, which in turn implies that all vector operations can be executed in constant time (as this is the case in the application and experimental evaluation). Otherwise, vector operations would introduce a further $O(\ell)$ factor to the total time complexity.

We note that if the zero vector does not belong to the set $\inputset$, then there are no vector spaces contained within it, and our problems have trivial solutions. Therefore, without loss of generality, we will implicitly assume that $\inputset$ always contains the zero vector.

\begin{remark}
    In the case $0\not\in\inputset$, set $\inputset$ can contain affine spaces, which are essentially translations of vector spaces~\cite{C81}. Our algorithms can also be adapted to the enumeration of affine spaces. 
    In fact, enumerating all affine spaces is equivalent to enumerating all vector spaces contained in the set ${\inputset}-u$, for all vectors $u\in {\inputset}$. The set ${\inputset}-u$ clearly contains the zero vector, and every vector space $\cal W$ enclosed in ${\inputset} -u$ corresponds to an affine space, ${\solution} +u$, enclosed in the original set $\inputset$. Note that for each vector $u\in {\cal S}$, both pre-processing and post-processing operations needed to reduce the enumeration of affine spaces to the enumeration of vector spaces require linear time.

\end{remark}

\subsection{Roadmap}
\label{section:map}
The rest of the paper is organized as follows. We start by approaching the problem(s) in a straightforward way, presenting a first brute-force algorithm in Section~\ref{section:brute-force}. Then, in Section~\ref{section:binpart}, we describe a more efficient enumeration algorithm based on binary partition, a widely employed paradigm in enumeration algorithms' development. Finally, Section~\ref{section:experiments} presents an experimental evaluation and comparison of our brute force and binary partition algorithms.

\section{A First Brute-force Algorithm}
\label{section:brute-force}
In this section, we present a brute force enumeration algorithm for the \textsc{Enclosed Space Enumeration} problem, which can be easily adapted to solve \textsc{Maximal Enclosed Space Enumeration} and \textsc{Maximum Enclosed Space} as well (more details in Section~\ref{subsec:bf-adaptation}). 

A brute-force algorithm consists in listing all potential solutions and checking, for each of them, whether it is actually a solution. 
In our case, a naive procedure could be to list all the \emph{subsets} $\solution$ of $\inputset$,
and for each such subset, verify whether it forms a vector space (closed under vector addition and scalar multiplication).

Still, listing all 
subsets would be costly, as the number of attempts would be exponential in $n = |\inputset|$.
Thanks to vector space properties, we can devise an algorithm that is still brute-force in nature, but significantly outperforms the naive procedure described above. 
The idea of the algorithm is to only list potential \emph{bases}\footnote{A basis is a set of vector through which any vector of the space can be expressed as a linear combination in a unique way (see Appendix~\ref{subsection:fields-vecspace} for more details).} $\mathcal{B}$ of solutions, and to then check whether $\text{Span}(\mathcal{B})$ is contained in $\inputset$. 
Since every basis $\mathcal{B}$ has cardinality at most $\log_2 n$, such a brute-force algorithm 
would perform a number of attempts on the order of $n^{\log_2 n}$, much more efficient than the previous naive approach, of order $2^n$.

We recall that we present our results over boolean vector spaces: $\bigspace$ is the set of binary vectors of length $\ell$, and the vector addition operation is given by the boolean XOR operation. The generalization of the brute-force approach to vector spaces over arbitrary finite fields \fc is described in Section~\ref{section:bf-generalization}. 

\subsection{The Algorithm}
\label{subsection:bf-algorithm}
A vector space of dimension $d$ over $\zdue$ has cardinality $2^d$: its elements are the binary $d$-tuples. 
Therefore, if $\solution$ is a space of dimension $d$ and is enclosed in $\inputset$, it must hold that $2^d \leq n$, that is, $d \leq \lfloor \log_2 n \rfloor$.
Thus, a possible algorithm consists of testing as candidate bases all possible subsets $\mathcal{B} \subseteq \inputset$ of cardinality at most $\lfloor \log_2 n \rfloor$.
To test whether such a subset is a valid solution, it suffices to  generate $\text{Span}(\mathcal{B})$, and check whether it is contained in $\inputset$.
If this is the case, then $\solution = \text{Span}(\mathcal{B})$ is a subspace of the desired type, and can be output as solution.

    Still, since each space can have multiple distinct bases, outputting every $\text{Span}(\mathcal{B})$ would generate duplicates. To avoid this issue, it suffices to assign to every space a unique, or canonical, basis. This can be easily done through, for instance, the well-known Gaussian-Jordan reduction algorithm (see Appendix~\ref{subsection:fields-vecspace} for more details). We can thus assume that we have a function $\textsc{is\_can\_basis}()$ which, given as input a set of vectors, checks in linear time whether the set forms a canonical basis.

In our algorithm, we thus proceed as follows (see Algorithm~\ref{algorithm:brute-force}):
we test all subsets $\mathcal{B}\subseteq \inputset$ of size at most $\lfloor \log_2 n \rfloor$, verifying first whether $\mathcal{B}$ is a canonical basis with function $\textsc{is\_can\_basis}(\mathcal{B})$; if so, we check whether the space $\solution = \text{Span}(\mathcal{B})$ is indeed a valid solution, using a second function $\textsc{is\_solution}(\solution,\inputset)$. For the \textsc{Enclosed Space Enumeration} problem, $\textsc{is\_solution}(\solution,\inputset)$ simply checks whether $\solution \subseteq \inputset$.

\begin{algorithm}
\caption{Brute-force enumeration of enclosed spaces}
\label{algorithm:brute-force}
\begin{algorithmic}[1]
\Procedure{Bruteforce\_enum}{$\inputset$}
    \State $n \gets |\inputset| $
    \State $b \gets \lfloor \log_2 n \rfloor$ \label{line:bf-cardinality-bound}
    \For{$\mathcal{B} \subseteq \inputset$ s.t. $|\mathcal{B}| \leq b$} \label{line:bf-forloop}
        \State $\solution \gets \text{Span}(\mathcal{B})$
        \If{$\textsc{is\_can\_basis}(\mathcal{B}) \ \land \ \textsc{is\_solution}(\solution,\inputset)$}
            \State Output $\solution$
        \EndIf
    \EndFor
\EndProcedure
\end{algorithmic}
\end{algorithm}

The correctness of the algorithm follows immediately from the previous discussion: by listing all the bases of size up to $\lfloor \log_2 n \rfloor$ we are only discarding subspaces that cannot be enclosed in \inputset; later, we explicitly check for them to be actual solutions, so every output set will surely be an enclosed space for \inputset.  The further check for canonicity of the basis ensures that we avoid duplication in the output, as every space is only considered once, when its canonical basis is chosen. 

\paragraph*{Complexity Analysis}
\label{section:bf-complexity}
In this section, we study the total-time complexity of the proposed brute-force enumeration of Algorithm~\ref{algorithm:brute-force}.

The sets over which the iterations are performed are the subsets of cardinality at most $\lfloor \log_2 n \rfloor$ of set \inputset of cardinality $n$, with one distinct set considered at each iteration.
For binomial coefficients $\binom{n}{i}$ such that $i=O(\log n)$, one can easily use the Stirling formula~\cite{knuth1989concrete} to show that $\binom{n}{i} = 2^{i\log_2 n - \Theta(i\log i)} $.

Thus, by summing all such set cardinalities we obtain the following number of iterations:
\begin{equation}
    \sum_{i=0}^{\lfloor \log_2 n \rfloor} \binom{n}{i} \le \sum_{i=0}^{\lfloor \log_2 n \rfloor} \binom{n}{\lfloor \log_2 n \rfloor} =  2^{\log_2^2 n - \Theta(\log n \log \log n)}, 
\end{equation}

The cost of each iteration is polynomial and, as such, is absorbed by the term $\Theta(\log n \log \log n)$ in our exponent, i.e., multiplying by a polynomial in $n$ corresponds to adding an $O(\log n)=o(\log n \log \log n)$ term to the exponent, which does not affect the previous expression. 
Indeed, checking whether a potential basis $\mathcal{B}$ is canonical (i.e. function \textsc{is\_can\_basis}) requires time proportional to the cardinality of the basis (see Appendix~\ref{subsection:fields-vecspace}), thus at most $O(\log_2 n)$. 
Furthermore, we can compute the Span of a set of $b = \lfloor \log_2(n) \rfloor$ vectors in $2^b\le n$ time, by trying all of their possible linear combinations.
Finally we can verify whether $\text{Span}(\mathcal{B})$ is included in $\inputset$ (i.e. function \textsc{is\_solution}) on the fly while computing $\text{Span}(\mathcal{B})$, with an overall cost of at most $O(n)$. 

As for the space, we only need to keep in memory \inputset and one of the potential solutions \solution at a time, and it is thus bounded by $O(n)$. 

We have thus arrived at the following result:

\begin{proposition}
\label{proposition:bruteforce}
Algorithm~\ref{algorithm:brute-force} enumerates all spaces enclosed in a set $\inputset$ of cardinality $n$ in total time $2^{\log_2^2 n - \Theta(\log n \log \log n)}$. The additional space required is $O(n)$.
\end{proposition}

One important downside of this enumeration algorithm is that its complexity is not output sensitive: when providing sets of the same size as input, it will always lead to the same total complexity regardless of the output size.
In Section~\ref{section:binpart} we will see a more refined algorithm which not only has faster total time complexity, but is also able to achieve linear delay, thus providing an output-sensitive running time as well. 

\subsection{Adaptation to Maximal and Maximum Versions}
\label{subsec:bf-adaptation}
In this section, we describe how to adapt Algorithm~\ref{algorithm:brute-force} to solve \textsc{Maximal Enclosed Space Enumeration} and \textsc{Maximum Enclosed Space} with the same total time complexity.

For \textsc{Maximal Enclosed Space Enumeration}, it suffices to change the \textsc{is\_solution} function accordingly, while still retaining the polynomial complexity.
A function that verifies the maximality of the vector space $\solution$ in $\inputset$ can be constructed by iterating over the elements $b \in \inputset \setminus \solution$ and checking that $\text{Span}(\mathcal{W},b) \not \subseteq \inputset$.
As for time complexity, this may require up to 
$|\solution| \cdot |\inputset \setminus \solution|$ 
time, that is, at most $O(n^2)$. Thus, we obtain an algorithm for this problem with total time and space complexity as given in Proposition~\ref{proposition:bruteforce}. 

For the \textsc{Maximum Enclosed Space} variant, in which one seeks only the maximum enclosed space, we can adopt the algorithm for \textsc{Enclosed Space Enumeration},
with an order modification and an early stopping condition. 
Indeed, in this case, it is important that the for-loop at line~\ref{line:bf-forloop} of Algorithm~\ref{algorithm:brute-force}, which iterates over all possible bases, does so in \emph{decreasing} order of cardinality: in this way, as soon as a valid space is found, the search can be stopped, and the found space can be returned as the solution of maximum cardinality. The enforcement of this order can be done at no additional cost with respect to the algorithm for \textsc{Enclosed Space Enumeration}, and thus we once again obtain an algorithm with the same total time complexity.

\subsection{Generalization to Finite Fields}
\label{section:bf-generalization}
The proposed brute-force algorithm can be generalized from boolean vector spaces to vector spaces \bigspace over any finite field \fc of cardinality $c$ as follows. 

The only dependence of the brute-force algorithm on the cardinality of the underlying field \zdue is given in line~\ref{line:bf-cardinality-bound}, while every other operation is completely independent from \zdue.
It now suffices to note that a vector space $\mathcal{W}$ of dimension $d$ over a field of cardinality $c$ has cardinality $|\mathcal{W}|=c^d$. 
Thus, similarly to before, to be contained in a given set $\inputset$ it is necessary for $\mathcal{W}$ to have dimension $d \leq \lfloor \log_c n \rfloor$. 
Consequently, the proposed brute-force algorithm 
can then be applied to any vector space $\bigspace$ over \fc in time $c^{\log_c^2 n - \Theta(\log n \log \log n)} = e^{\frac{1}{\ln c}\ln^2 n - \Theta(\log n \log \log n)}$. The space complexity remains unchanged. 

\section{Binary Partition Approach}
\label{section:binpart}

Binary partition is a recursive technique for enumerating objects. 
The recursive function $\textsc{rec\_enum}(\solution,\mathcal{X},\inputset)$ considers a partition of $\inputset$ into three sets: $\solution$, the set of elements that must necessarily be in the desired subset; $\mathcal{X}$, the set of elements that must necessarily be excluded from the desired subset; and  $\mathcal{R} = \inputset \setminus (\solution \cup \mathcal{X})$, the remaining elements to consider. 
The paradigm works as follows: at each step, the algorithm considers an unprocessed element $r\in \mathcal{R}$ and tries to add it (separately) to both $\solution$ and $\mathcal{X}$, partitioning the solution space into solutions containing $r$, and solutions that do not contain $r$. It is fairly trivial to observe how the algorithm finds all solutions.
That is, the binary partition algorithm takes an element $r \in \mathcal{R}$ and recursively calls $\textsc{rec\_enum}({\solution} \cup \{r\}, \mathcal{X}, \inputset)$ and $\textsc{rec\_enum}({\solution}, \mathcal{X} \cup \{r\}, \inputset)$. Once $\mathcal{R}$ is empty, the result is output if $\solution$ 
is actually a solution.\footnote{We output only when $\mathcal{R}$ is empty, i.e., on leaves, to avoid multiple outputs of the same solution, since $\textsc{rec\_enum}({\solution}, \mathcal{X} \cup \{r\}, \inputset)$ does not change $\solution$.}

The algorithm can be represented as a binary tree of calls to $\textsc{rec\_enum}$, such that the overall complexity is the complexity of a single call to $\textsc{rec\_enum}$ multiplied by the number of calls.
Usually, it is possible to prune the tree by adding a $\textsc{can\_be\_extended}(r,\solution,\mathcal{X},\inputset)$ function that returns $\texttt{true}$ if it exists a solution $\mathcal{P}$ of the problem such that ${\solution}\cup \{r\} \subseteq \mathcal{P} \subseteq \inputset\setminus \mathcal{X}$, and $\texttt{false}$ otherwise. Then, it would suffice to perform the two recursive calls for $r$ only when the output of such function is \texttt{true}.
In this way, we don't need to explicitly check for $\solution$  to be a solution once $\mathcal{R}=\emptyset$, as we avoid exploring branches that don't lead to any solution. 
Furthermore, whenever $\textsc{can\_be\_extended}(r,\solution,\mathcal{X},\inputset)$ is $\texttt{false}$,
it is guaranteed that no solution found in the subtree $\textsc{rec\_enum}(\solution, \mathcal{X}, \inputset)$ will contain $r$, so we can add $r$ to $\mathcal{X}$ without any consequences. Another upside of this choice is that we can skip $r$ in further iterations among $\mathcal{R}$, thus improving efficiency. Finally, as $\texttt{false}$ implies a single recursive child, we may instead iterate among the elements $r\in \mathcal{R}$ until an $r$ such that $\textsc{can\_be\_extended}(r,\solution,\mathcal{X},\inputset)$ is $\texttt{true}$ is found, or until we get to the end of $\mathcal{R}$.
These tree pruning strategies are what make the binary partition algorithm faster than a simple brute-force enumeration over all subsets.

\subsection{A Binary Partition Algorithm for Enclosed Space Enumeration}
Algorithm~\ref{algorithm:binarypartition} shows the pseudocode of our binary partition algorithm for \textsc{Enclosed Space Enumeration} problem, following the paradigm described in the previous section. 

We first need to appropriately define function $\textsc{can\_be\_extended}(r,\solution,\mathcal{X},\inputset)$. 
In our case, it must simply verify whether $\text{Span}({\solution}\cup \{r\})\subseteq \inputset \setminus \mathcal{X}$, as otherwise it would be impossible to extend ${\solution} \cup \{r\}$ to a subspace of $\bigspace$ contained in $\inputset\setminus \mathcal{X}$. 

However, the powerful aspect is that all the elements of $\text{Span}(\solution\cup \{r\})\setminus \text{Span}(\solution)$ are equivalent: if a solution contains $\solution$ and $r$, it must contain all elements of the form $s+\alpha\cdot r$ with $s \in \solution$ and $\alpha$ in the field relative to our vector space (in our case, $\zdue$); similarly, if a solution contains $\solution$ and any element of the form $s+\alpha\cdot r$ with $s \in \solution$ and $\alpha \neq 0$, it must contain the element $-\alpha^{-1}s + \alpha^{-1}(s+\alpha\cdot r)=r$. 
This allows us to consider all the elements of $\text{Span}(\solution\cup \{r\})\setminus \text{Span}(\solution)$ in a single step, i.e., when we add $r$ we can immediately add them all at the same time to $\solution$ or $\mathcal{X}$ without loss of generality (as we would be forced to do so anyway in the recursive subtree).

If, at each step of the extension process, we extend $\solution$ with the whole $\text{Span}(\solution\cup \{r\})\setminus \text{Span}(\solution)$ set, it can be shown by induction that $\solution$ is a vector space at every step.
This way, $\text{Span}(\solution\cup \{r\}) = \solution \cup (r+\solution)$, where $r+\solution = \{r + w \ | \ w \in \solution \}$. 
The two recursive calls will thus be $\textsc{rec\_enum}(\solution\cup (r+\solution),\mathcal{X},\inputset)$ and $\textsc{rec\_enum}(\solution,\mathcal{X}\cup (r+\solution),\inputset)$.

\begin{algorithm}
\caption{Binary Partition for Enclosed Space Enumeration}\label{algorithm:binarypartition}

\begin{algorithmic}[1]
    \State Call $\textsc{rec\_enum}(\{0\},\emptyset,{\inputset} \setminus \{0\})$

\Procedure{rec\_enum}{$\solution, \mathcal{X}, \inputset$}
    
    \If{$\exists r \in \inputset \setminus (\solution\cup \mathcal{X})$ s.t. $\textsc{can\_be\_extended}(r,\solution,\mathcal{X},\inputset)$} \label{line:bp-exists-extension}
        \State Call $\textsc{rec\_enum}(\solution\cup (r+\solution),\mathcal{X},\inputset)$
        \State Call $\textsc{rec\_enum}(\solution,\mathcal{X}\cup (r+\solution),\inputset)$
    \ElsIf{$\textsc{is\_solution}(\solution,\inputset)$}
        \State Output $\solution$
    \EndIf
\EndProcedure
\end{algorithmic}
\end{algorithm}

The function $\textsc{can\_be\_extended}(r,\solution,\mathcal{X},\inputset)$ can be implemented as follows: it iterates over the elements $s \in \solution$ and checks whether $r + s \in \inputset$. The various $r + s$ values obtained in this way (including $r$ itself) can be accumulated in a set $\mathcal{L}$. If all the $r + s$ elements belong to $\inputset$, then $\textsc{can\_be\_extended}(r,\solution,\mathcal{X},\inputset)$ outputs \texttt{true} and we have already computed $\mathcal{L} = r + \solution$, saving additional computations in the recursive calls. On the other hand, if any $r + s \not\in \inputset$ is found, then $\textsc{can\_be\_extended}(r,\solution,\mathcal{X},\inputset)$ outputs \texttt{false}, and as a further optimization
the set $\mathcal{L}$ computed so far can be moved from $\mathcal{R}$ to $\mathcal{X}$, so that each of its elements is never computed more than once.

\begin{remark}\label{rmk:root-to-leaf}
    Observe how each element of $\mathcal{R}$ is considered only once not only in a specific recursive call, but in any root-to-leaf path, as whenever it is considered it is added to either $\mathcal{X}$ or $\solution$.
\end{remark}

Correctness of the proposed algorithm thus follows from the definition of the $\textsc{can\_be\_extended}$ function, together with the correctness of the binary partition paradigm.

Note that, since in our algorithm it is ensured by construction that {\solution} is a vector space and that $\solution\subseteq \inputset$, $\textsc{is\_solution}(\solution,\inputset)$ 
can be chosen to always return \texttt{true}.

\begin{remark}
  \label{rmk:iterative-binpart}
    We remark that it is of course possible to implement an equivalent iterative algorithm, which typically improves space usage as it stores just the differences between one simulated recursive call and the nested one, and may obtain a practical speedup by avoiding the recursion stack.
\end{remark}

\paragraph*{Complexity Analysis}
Let us now study the complexity of the binary partition enumeration algorithm described in the previous section. 

As mentioned before, the total time complexity is given by the time required by a single call to $\textsc{rec\_enum}$, multiplied by the number of such calls, that is, the number of nodes in the recursion tree.  

In a regular binary partition, the depth of the subtree corresponding to call $\textsc{rec\_enum}(\solution,\mathcal{X},\inputset)$ is, in the worst case, $|\mathcal{R}|$, since at each level one element of $\mathcal{R}= \inputset \setminus (\solution \cup \mathcal{X})$ is considered. However, in our algorithm the depth is at most $\frac{|\mathcal{R}|}{|\solution|}$, since at each level all the elements of $r+\solution$ are considered (included or excluded) simultaneously, and $|r+\solution|=|\solution|$.

Let $f(m)$ be the maximum number of leaves that a recursion tree of depth $m$ can have for our problem. Naturally, we have $f(0)=1$.
In the general case, the number of leaves of the subtree corresponding to the call $\textsc{rec\_enum}(\solution,\mathcal{X},\inputset)$ is the sum of the number of leaves of the two recursive calls $\textsc{rec\_enum}(\solution,\mathcal{X}\cup (r+\solution),\inputset)$ and $\textsc{rec\_enum}(\solution\cup (r+\solution),\mathcal{X},\inputset)$.
If the depth of the subtree corresponding to the call $\textsc{rec\_enum}(\solution,\mathcal{X},\inputset)$ is $m = \left\lfloor\frac{|\mathcal{R}|}{|\solution|}\right\rfloor$, then the depth of the subtree corresponding to $\textsc{rec\_enum}(\solution,\mathcal{X}\cup (r+\solution),\inputset)$ is $\left\lfloor\frac{|\mathcal{R}|-|\solution|}{|\solution|} \right\rfloor = m-1$, since $|\solution|$ elements are removed from $\mathcal{R}$ while the size of $\solution$ stays the same. At the same time, the depth of the subtree corresponding to $\textsc{rec\_enum}(\solution\cup (r+\solution),\mathcal{X},\inputset)$ is $\left\lfloor\frac{|\mathcal{R}|-|\solution|}{2\cdot|\solution|} \right\rfloor = \left\lfloor \frac{m-1}{2}\right\rfloor$, since $|\solution|$ elements are removed from $\mathcal{R}$, while the size of $\solution$ is increased with the same $|\solution|$ elements removed from $\mathcal{R}$.

Our function $f$ therefore satisfies the following recurrence relation:
\begin{equation}
\label{eq:recurrence}
    f(m) = f(m-1) + f\left(\left\lfloor \frac{m-1}{2}\right\rfloor\right).
\end{equation}
It is proven in the literature~\cite{deBruijn1948} that this equation has solution:
\begin{equation}
\label{eq:fn}
    f(n) = 2^{\frac{1}{2}\log_2^2 n-\Theta(\log n \log\log n)}.
\end{equation}

Since all recursive call, other than the leaves, will have at least 2 recursive children, the recursion tree has no unary nodes, and it is known that in such a tree the total number of nodes is at most twice the number of leaves, so it can be bounded as well by $2^{\frac{1}{2}\log_2^2 n-\Theta(\log n \log\log n)}$.

To conclude, we need to bound the amount of computation performed per recursive call. 
If the function $\textsc{can\_be\_extended}(r,\solution,\mathcal{X},\inputset)$ is implemented as previously described, the time complexity of 
line~\ref{line:bp-exists-extension}
is at worst $O(n)$. 
Since \textsc{is\_solution} can be considered to be always true, and thus requires no computation, each call to $\textsc{rec\_enum}(\solution,\mathcal{X},\inputset)$ requires at most $O(n)$ time.
This cost can be absorbed by the $-\Theta(\log n \log\log n)$ term in the exponent of $f(n)$, as already described in \ref{section:bf-complexity}.
We can now state the following result:

\begin{proposition}
\label{proposition:binarypartition}
Algorithm~\ref{algorithm:binarypartition} enumerates all spaces enclosed in a set $\inputset$ of cardinality $n$ in total time $2^{\frac{1}{2}\log_2^2 n - \Theta(\log n \log \log n)}$.
\end{proposition}
This algorithm is indeed a relevant speedup, as its time complexity is the square root of that of the brute-force approach.

Another important feature of this solution compared to the brute-force one is that it
is output-sensitive: this means that computing the spaces enclosed in a set is faster for sets with fewer enclosed spaces. 
In particular, consider any root-to-leaf path. The total number of operations performed across all calls to $\textsc{rec\_enum}$ along this path is $O(n)$: this is because, as per Remark~\ref{rmk:root-to-leaf}, each element of $\inputset$ is processed exactly once and added to either $\solution$ or $\mathcal{X}$. The same applies during the backtracking phase, where the sets $\solution$ and $\mathcal{X}$ are emptied. 

Since a solution is output at each leaf, and the path between one leaf and the next consists only of an ascending phase and a descending phase, we have proved the following result:

\begin{proposition}
\label{proposition:binarypartition_delay}
Algorithm~\ref{algorithm:binarypartition} can enumerate all spaces enclosed in a set $\inputset$ of cardinality $n$ with delay $O(n)$.
\end{proposition}

Finally, let us analyze the space complexity. The memory complexity of a single recursive call $\textsc{rec\_enum}(\solution,\mathcal{X},\inputset)$ is $O(n)$, which, multiplied by the depth of the recursion tree (also $O(n)$), gives an overall space complexity of $O(n^2)$ for the recursive implementation.

The iterative version described in Remark~\ref{rmk:iterative-binpart}, on the other hand, needs to only store the differences in $\solution$ and $\mathcal{X}$ between each level and the previous one. Since all these differences are disjoint, the memory used by their union is at most $O(n)$. This allows us to conclude that:

\begin{proposition}
\label{proposition:binarypartition_space}
Algorithm~\ref{algorithm:binarypartition} can enumerate all spaces enclosed in a set $\inputset$ of cardinality $n$ with total additional space $O(n)$.
\end{proposition}

\subsection{Adaptation to Maximal and Maximum Versions}
\label{subsec:bp-adaptation}
In this section, we describe how to adapt Algorithm~\ref{algorithm:binarypartition} to solve \textsc{Maximal Enclosed Space Enumeration} and \textsc{Maximum Enclosed Space} with the same total time complexity.

For \textsc{Maximal Enclosed Space Enumeration}, it suffices to change the \textsc{is\_solution} function to check for maximality, which can be done in $O(n^2)$ time as described in Section~\ref{subsec:bf-adaptation}.

As for the overall time and memory, we obtain an algorithm for this problem with complexities as given in Propositions~\ref{proposition:binarypartition} and~\ref{proposition:binarypartition_space}. 

Unfortunately, there is no result equivalent to Proposition~\ref{proposition:binarypartition_delay}. This issue, which is common when using binary partition to find maximal solutions, is due to the possibility of traversing entire recursive subtrees in which all the solutions found are strict subsets of other solutions located in different parts of the tree.

Regarding the \textsc{Maximum Enclosed Space} problem, one possible approach is to enumerate the solutions found using Algorithm~\ref{algorithm:binarypartition} while keeping in memory the largest solution $\mathcal{M}$ found so far.
However, it is important to note that the maximum enclosed space obtainable through a call to $\textsc{rec\_enum}(\solution, \mathcal{X}, \inputset)$ must be contained in $\inputset \setminus \mathcal{X}$. If $\mathcal{X}$ is sufficiently large, $\inputset \setminus \mathcal{X}$ might be too small to contain a solution larger than $\mathcal{M}$, making the call to $\textsc{rec\_enum}(\solution, \mathcal{X}, \inputset)$ unnecessary.

For this reason, we can improve the algorithm by adding a check at the beginning of $\textsc{rec\_enum}(\solution, \mathcal{X}, \inputset)$ that compares the cardinalities of $\inputset \setminus \mathcal{X}$ and $\mathcal{M}$, potentially pruning the entire subtree rooted at $\textsc{rec\_enum}(\solution, \mathcal{X}, \inputset)$ in a branch-and-bound fashion.

Naturally, for this optimization to be effective, it is crucial that the branch in which the solution is extended—namely, $\textsc{rec\_enum}(\solution \cup (r + \solution), \mathcal{X}, \inputset)$—is always explored first.
This check doesn't provide any additional cost with respect to the previous algorithm for \textsc{Enclosed Space Enumeration}, and thus we once again obtain an algorithm with the same total time complexity.

\subsection{Generalization to Finite Fields}
\label{section:bp-generalization}

Algorithm~\ref{algorithm:binarypartition} generalizes easily to finding enclosed spaces $\mathcal{W}$ over a field of finite cardinality $c$: indeed, we only need to replace $r+\solution$ with $\text{Span}(\solution\cup \{r\}) \setminus \solution$.

Since in a finite field of cardinality $c$ it holds that $|\text{Span}(\solution\cup \{r\})| = c \cdot|\solution|$, the recurrence of eq.~\eqref{eq:recurrence} becomes:
\begin{equation}
    f(m) = f(m-1) + f\left(\left\lfloor \frac{m-1}{c}\right\rfloor\right).
\end{equation}

The solution is again known in the literature~\cite{deBruijn1948}, and is given by:
\begin{equation}
\label{eq:fn-finitefield}
    f(n) = e^{\frac{1}{2\ln c}\ln^2 n-\Theta(\log n \log\log n)}.
\end{equation}
By repeating the same considerations of the previous complexity analysis, the resulting algorithms have total time complexity given by $c^{\frac{1}{2}\log_c^2 n-\Theta(\log n \log\log n)} = e^{\frac{1}{2\ln c}\ln^2 n-\Theta(\log n \log\log n)}$, while delay and space are unchanged.

\section{Experimental Results}
\label{section:experiments}
We now present the experimental evaluation of our algorithms. 

As the problem is motivated by the field $\mathbb{Z}/(2)$, i.e., binary strings, and these allow further practical optimizations, our code was written specifically for them. 

All tests consider sets of binary strings randomly generated with a fixed length between $4$ and $16$. Indeed, given the length of these strings, it was possible to store them as atomic \texttt{usize} values, leveraging Rust's native bitwise operations for efficient manipulation and computation.
This approach allowed us to keep the representation compact while taking advantage of fast, low-level operations.\footnote{Nonetheless, one can expect an implementation for generic finite fields to be slower by just the constant factors required for the slower string operations.}

All tests were conducted on a DELL PowerEdge R750 machine in a non-exclusive mode. This platform features 24 cores with 2 Intel(R) Xeon(R) Gold 5318Y CPUs at 2.10 GHz, and 976 GB of RAM. The operating system is Ubuntu 22.04.2 LTS.

\begin{figure}[h!]
    \centering
    \begin{subfigure}[b]{0.32\textwidth}
        \includegraphics[width=\textwidth]{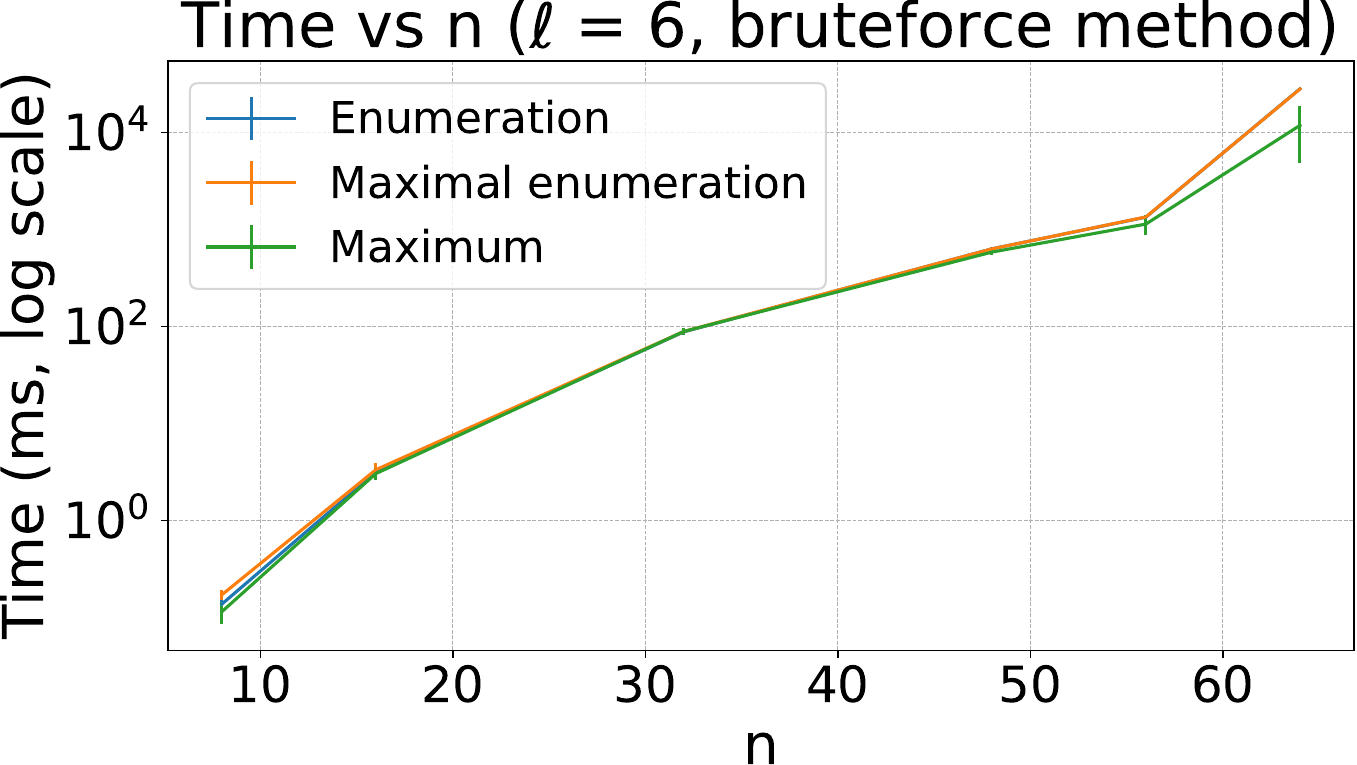}
        \caption{Brute-force}
    \end{subfigure}
    \begin{subfigure}[b]{0.33\textwidth}
        \includegraphics[width=\textwidth]{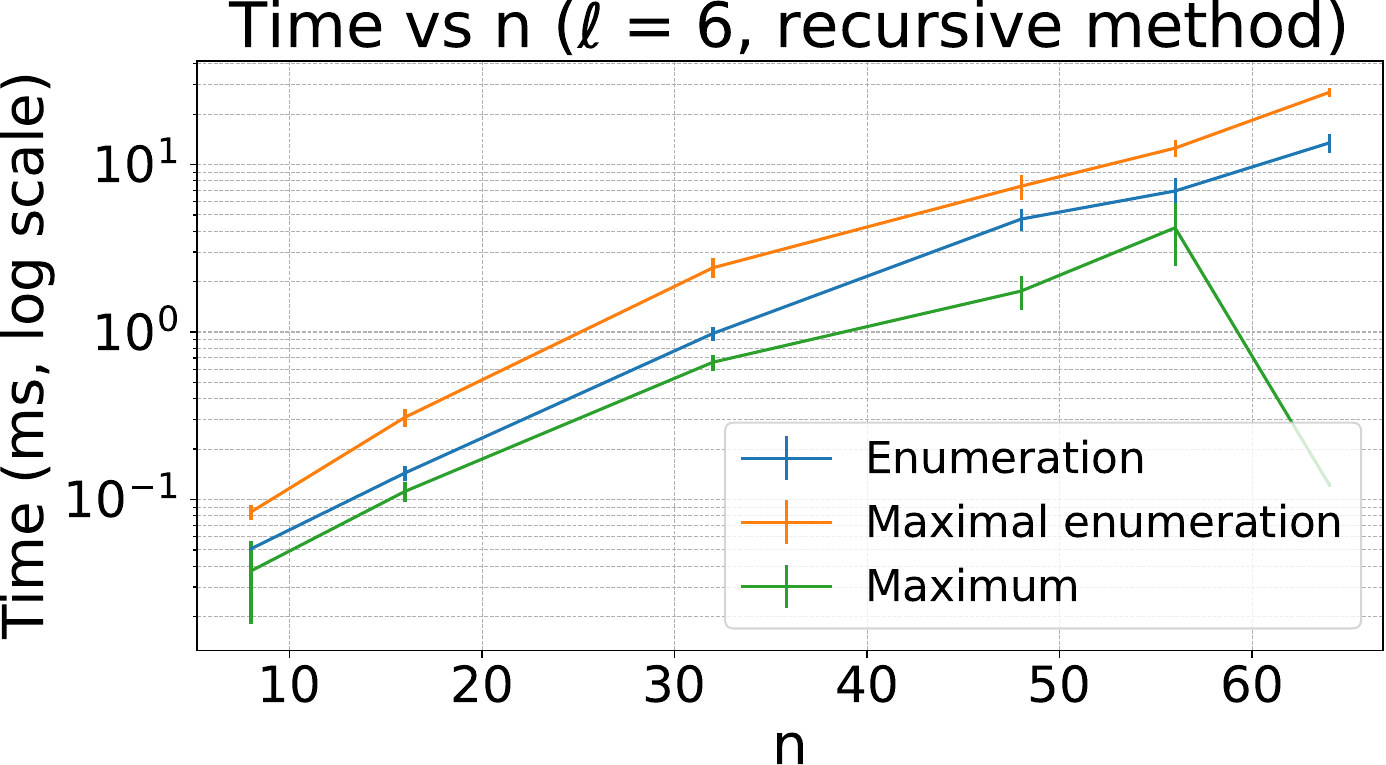}
        \caption{Recursive binary partition}
    \end{subfigure}
    \begin{subfigure}[b]{0.33\textwidth}
        \includegraphics[width=\textwidth]{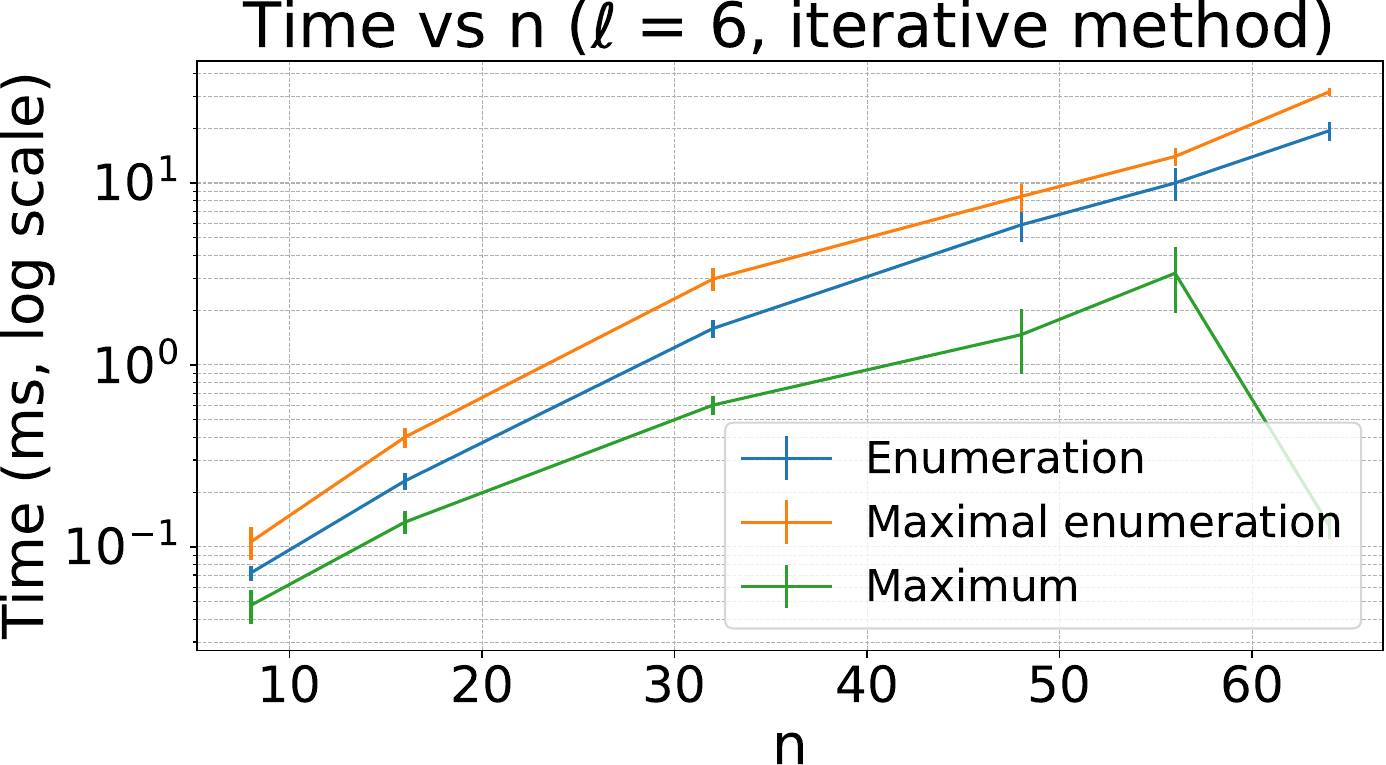}
        \caption{Iterative binary partition}
    \end{subfigure}
    
    \caption{Running time (ms, log scale) for the 3 problems considered, using the (a) Brute-force algorithm, (b) recursive Binary Partition, and (c) iterative Binary Partition, for varying number $n$ of vectors of length $\ell=6$.}
    \label{fig:tasks_comparison}
\end{figure}

Our implementations are sequential and written in Rust. The source code will be available in the final, non anonymized, version. 
We measured both runtime and memory usage, comparing the recursive and iterative versions of our binary partition algorithm 
with the brute-force one, to highlight the theoretical and practical advantages discussed in the previous sections. The computational times are shown in Figure~\ref{fig:tasks_comparison}.

In what follows, we report aggregate statistics over 100 runs per configuration, whenever the run took less then $1 \text{ min}$, or 10 runs per configuration, whenever the run took between $1 \text{ min}$ and $1 \text{ h}$.

As expected, the algorithm for solving \textsc{Maximal Enclosed Space Enumeration} is uniformly slower, since it performs the same computational steps as \textsc{Enclosed Space Enumeration} plus additional operations in the leaves. The slowdown appears proportionally less significant in the brute-force approach, given the higher total time of the execution.

The branch-and-bound strategy used in \textsc{Maximum Enclosed Space} provides noticeable, albeit not dramatic, speedup compared to enumeration. The most significant difference naturally arises in the case where the set $\inputset$ coincides with the entire space $\bigspace$, as the algorithms quickly find a solution of the same size as the input, which allows immediate termination.

Since their asymptotic complexity is the same, the rest of the experiments will show only the \textsc{Enclosed Space Enumeration} variant.

\begin{figure}[h!]
    \begin{subfigure}[b]{0.32\textwidth}
        \includegraphics[width=\textwidth]{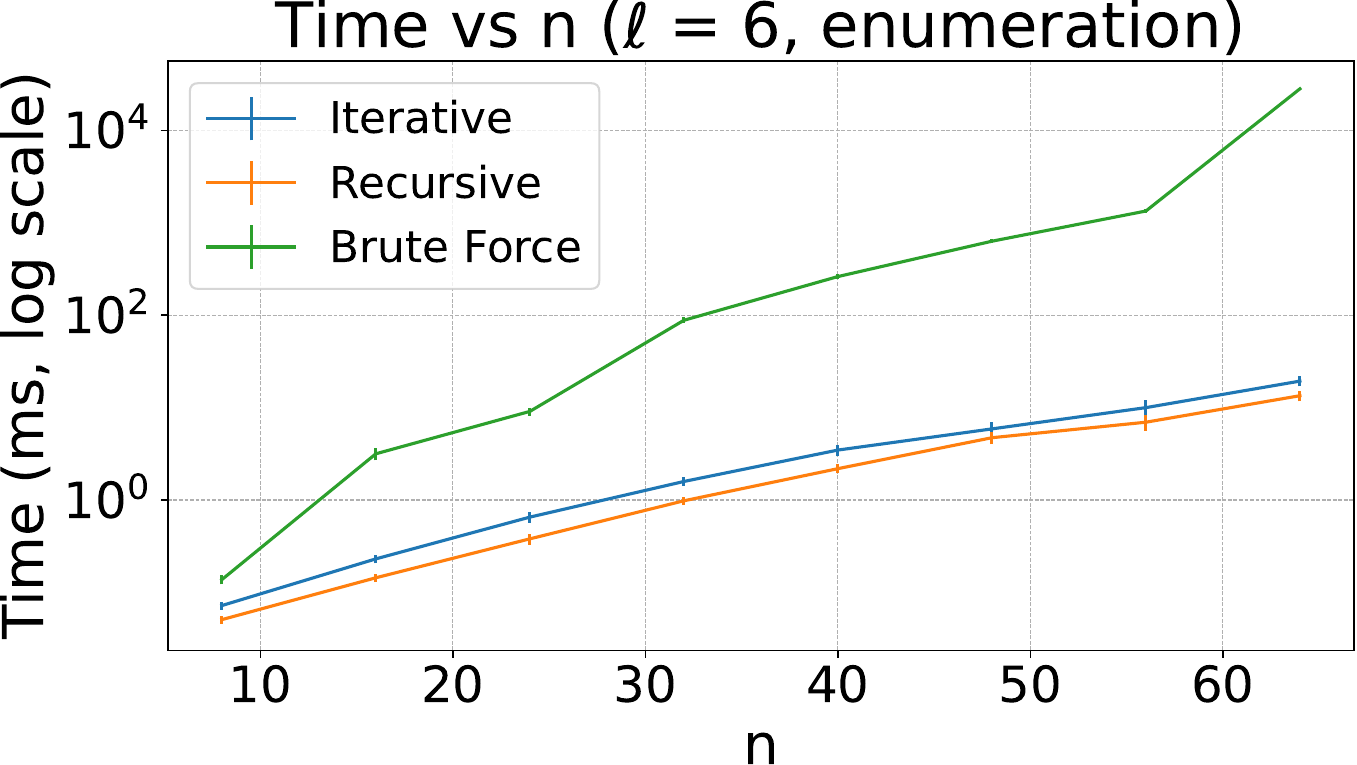}
        \caption{Comparison for $\ell=6$}
    \end{subfigure}
    \begin{subfigure}[b]{0.32\textwidth}
        \includegraphics[width=\textwidth]{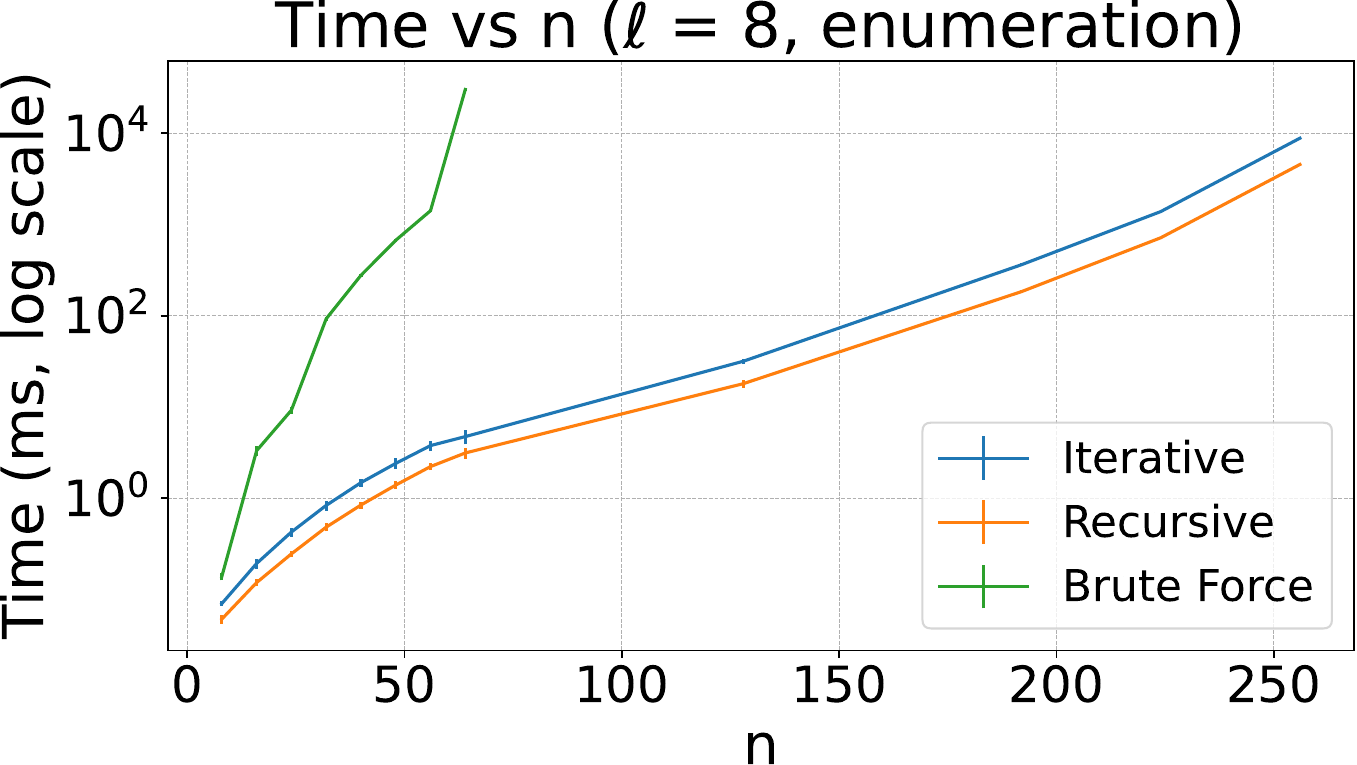}
        \caption{Comparison for $\ell=8$}
    \end{subfigure}
    \begin{subfigure}[b]{0.32\textwidth}
        \includegraphics[width=\textwidth]{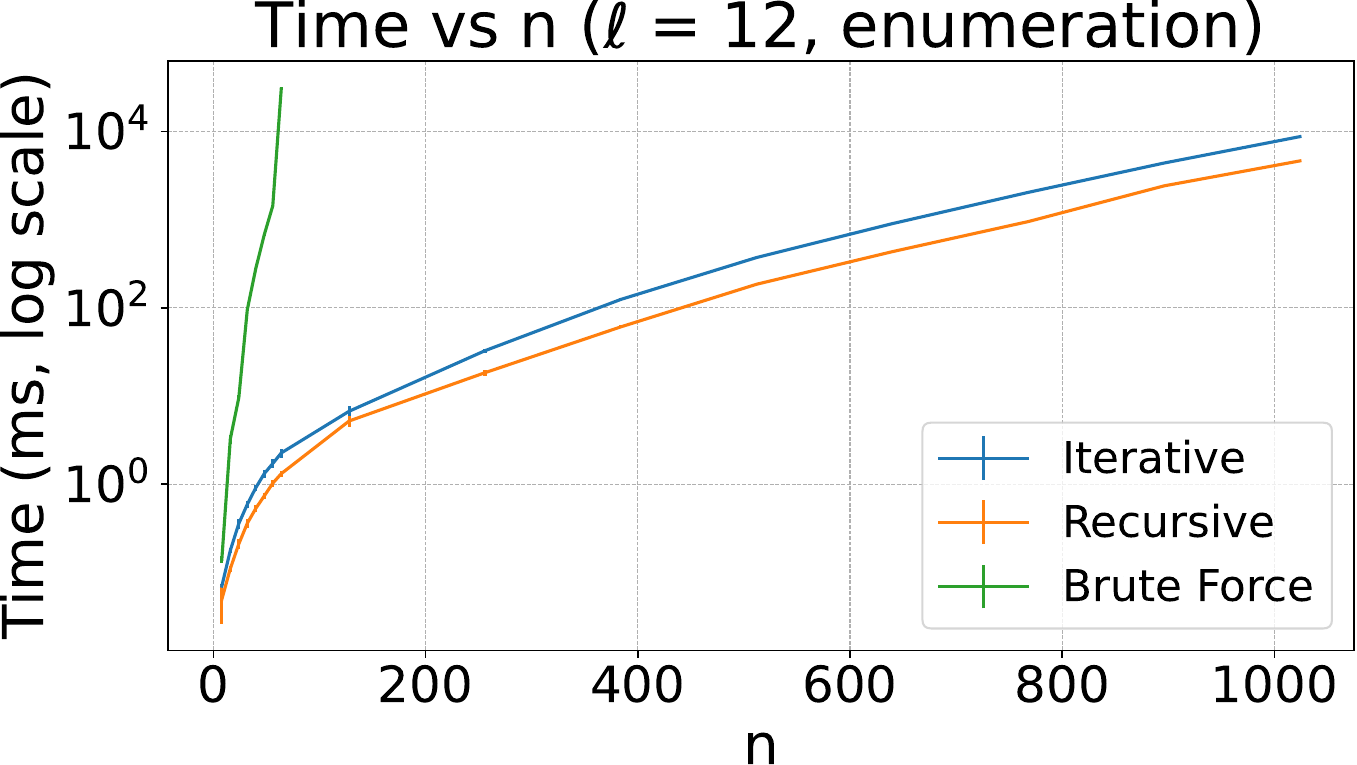}
        \caption{Comparison for $\ell=12$}
    \end{subfigure}
    
    \caption{Running time (ms, log scale) comparison for the \textsc{Enclosed Space Enumeration} problem between the brute-force algorithm, recursive binary partition, and iterative binary partition, for varying number $n$ of vectors of length $\ell=6,8,12$.}
\label{fig:methods_comparison}
\end{figure}

Figure~\ref{fig:methods_comparison} shows a running time comparison of the 3 methods for different values of the dimension $\ell$ of the space \bigspace. The brute-force approach is clearly slower: for $\ell=8,12$ it hits the timeout for $n$ in the tens, while the binary partition methods remain manageable for inputs several orders of magnitude larger. The iterative version is consistently slightly slower, which is probably due to the upkeeping overhead for maintaining $\solution$ rather than simply copying and deleting it. Other than this marginal difference, as expected the behaviour of the two variants is almost identical, with growth curves noticeably less sloped than the brute-force approach, which suggests an exponentially smaller growth.

\begin{figure}[h!]
    \begin{subfigure}[b]{0.48\textwidth}
        \includegraphics[width=\textwidth]{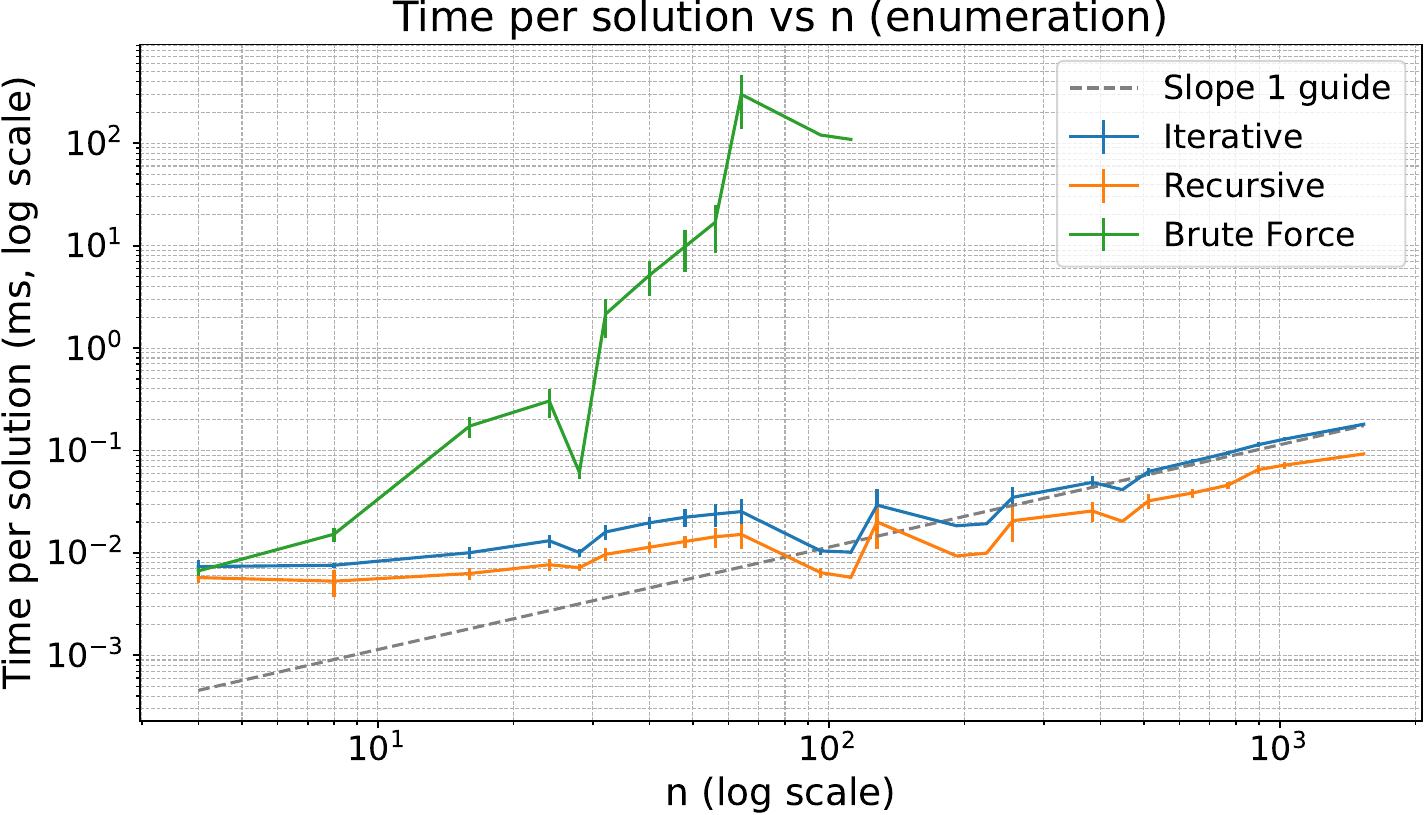}
    \caption{\textsc{Enclosed Space Enumeration}}
    \end{subfigure}
    \begin{subfigure}[b]{0.48\textwidth}
        \includegraphics[width=\textwidth]{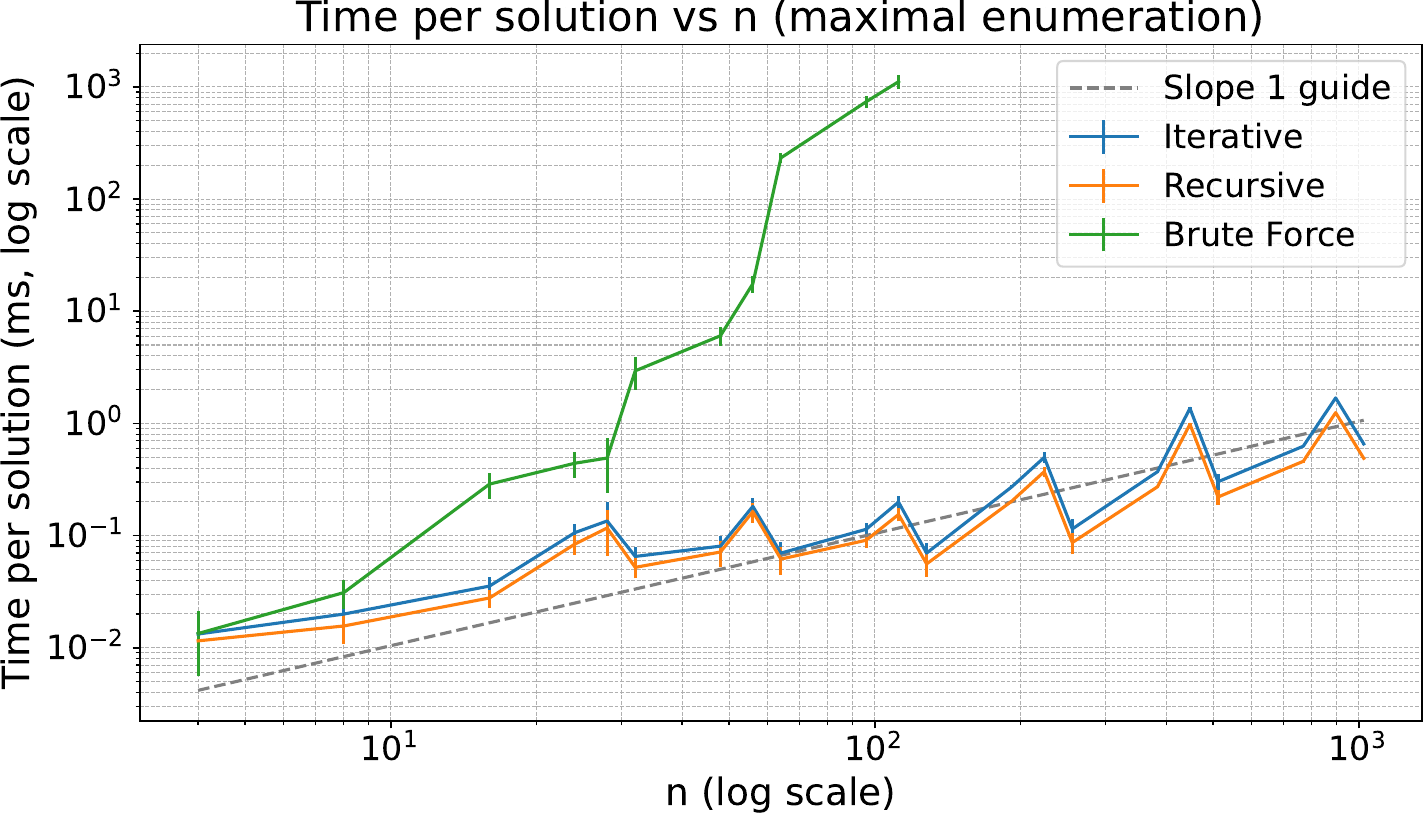}
    \caption{\textsc{Maximal Enclosed Space Enumeration}}
    \end{subfigure}
    \caption{
    Running time per solution (ms, log scale) comparison for the (a) \textsc{Enclosed Space Enumeration} problem, and (b) \textsc{Maximal Enclosed Space Enumeration} problem, between the brute-force algorithm, recursive binary partition, and iterative binary partition, for varying number $n$ of vectors.
    }
    \label{fig:output_sensitivity}
\end{figure}

The plots of Figure~\ref{fig:output_sensitivity} show instead the relationship between the size of the input ($n$) and the average time required to find each solution (total time divided by number of solutions).

The data highlights the output-sensitive behavior of the binary partition method, in the case of \textsc{Enclosed Space Enumeration}: in particular, the points are distributed around a line with slope $1$ in the log-log plot, indicating a direct proportionality between the input size and the execution time per solution.
This is consistent with the theoretical results, since the average time per solution equals the average delay between consecutive solutions.

Remarkably, although with greater variability, the same can be said for \textsc{Maximal Enclosed Space Enumeration}, despite no theoretical output-sensitive guarantee.

Looking at the brute-force approach, we see that not only the total time, but even the time per solution grows dramatically with the size of the problem, suggesting that -unsurprisingly- this approach becomes progressively less efficient as the problem size gets bigger.

The lack of output sensitivity of the brute-force algorithm is made even clearer in the 
plot in Figure~\ref{fig:density}, that refers to the case where $n=64$.

\medskip 

\begin{minipage}[c]{.5\textwidth}
    \centering
    \includegraphics[width=.8\textwidth]{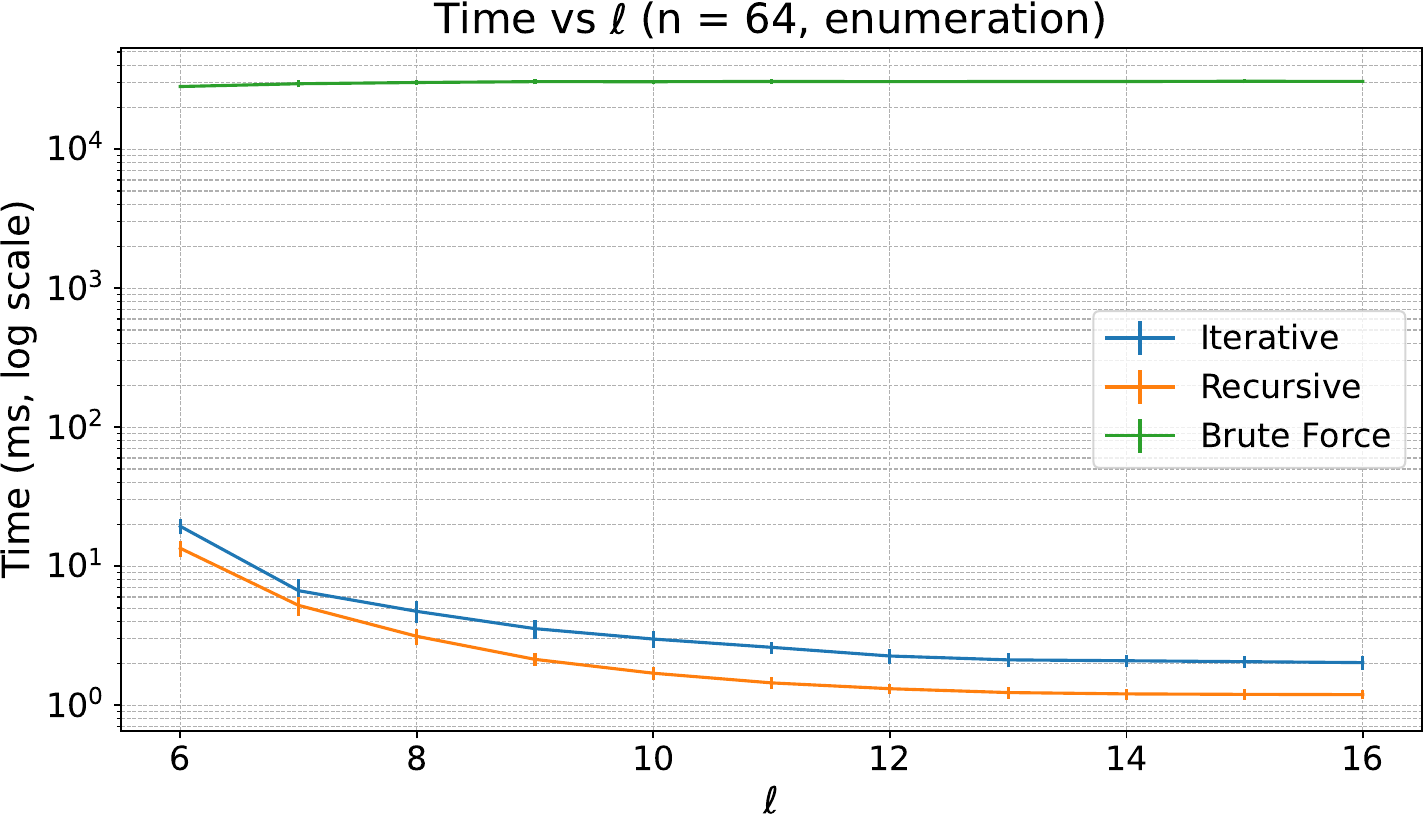}
    \captionof{figure}{
    Running time (ms, log scale) comparison for the enumeration problem between the brute-force algorithm, recursive binary partition, and iterative binary partition, for varying vector lengths $\ell$, with input size $n = 64$.}
    \label{fig:density}
\end{minipage}
\hfill
\begin{minipage}[c]{.4\textwidth}
    \centering
    \includegraphics[width=\textwidth]{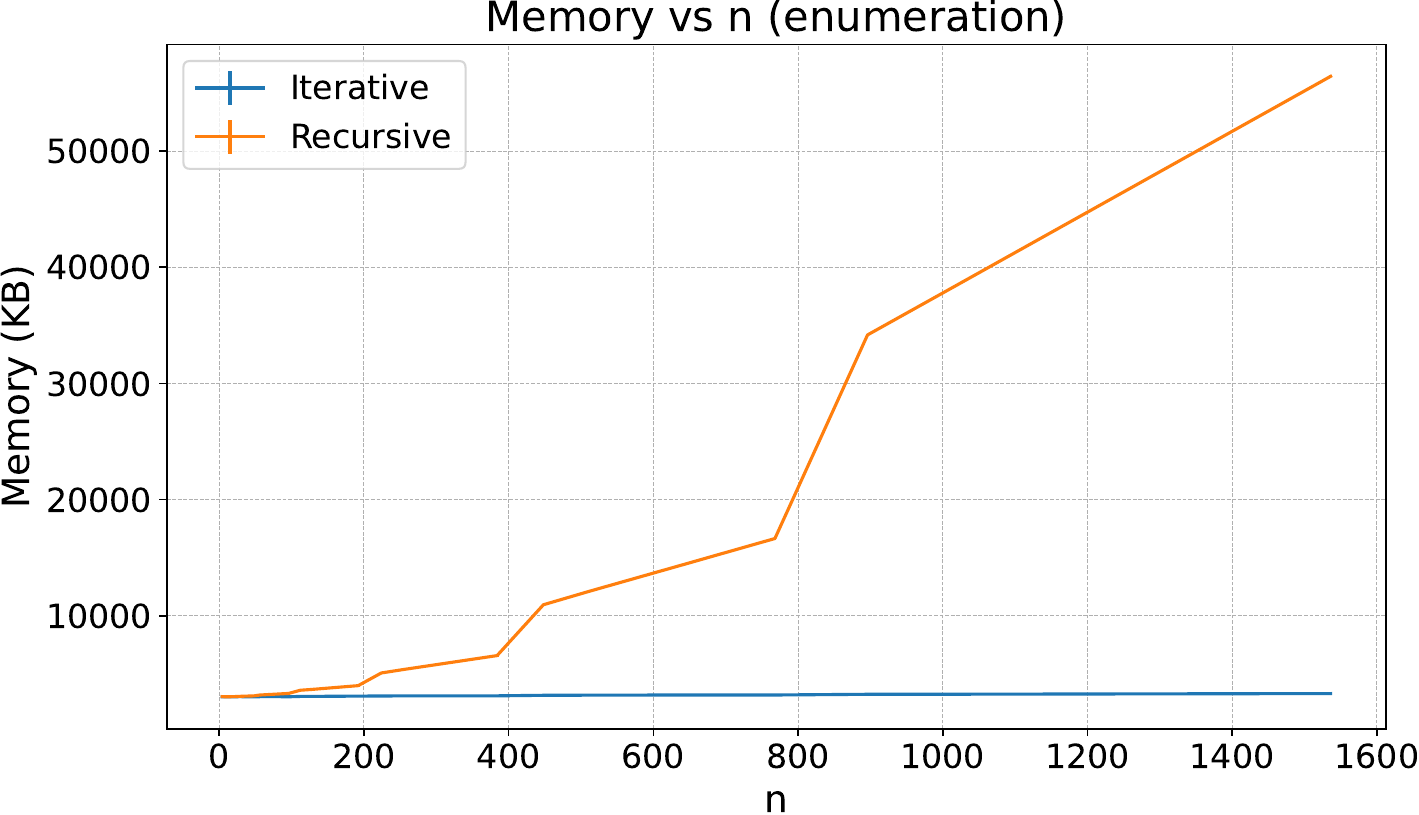}
    \captionof{figure}{
    Memory (KB) comparison for the enumeration problem between the recursive binary partition and iterative binary partition for varying number $n$ of vectors.}
    \label{fig:memory}
\end{minipage}

\medskip

For fixed $n$, increasing $\ell$ also increases the sparsity of the vectors and thus decreases the number of solutions (see Appendix~\ref{section:study_on_spaces}). This leads to a decrease in execution time for the binary partition method, while it makes no difference for the brute-force approach.

We conclude with the plot of Figure~\ref{fig:memory}, which confirms the lower memory usage of the iterative version: given the modest values of $n$, the expected $O(n)$ memory consumption becomes negligible compared to noise in the iterative case, where the observed memory usage appears effectively constant. In contrast, the increase in memory usage is clearly visible in the recursive version, where the expected consumption is $O(n^2)$.

\section{Conclusion}

In this paper we have designed and experimentally evaluated   an  algorithm for the enumeration of (maximal) vector spaces over finite fields enclosed in a given set, motivated by the search of structural regularities in Boolean functions. The proposed algorithm is based on the binary partition paradigm and has a total time complexity of  $e^{\frac{1}{2\ln c}\ln^2 n - \Theta(\log n \log \log n)}$, where $n$ is the number of vector and $c$ is the finite field cardinality.  Compared to a  brute-force approach, the algorithm  provides a speed-up that is quadratic in theory, which appears even more significant in the experimental evaluation.

Interesting future directions are refining the enumeration of only maximal/maximum enclosed spaces, and the generalization from vector spaces to different algebraic structures.

\bibliography{spp}

\appendix

\section{Notions of Fields and Vector Spaces }
\label{subsection:fields-vecspace}
This paper deals with vector spaces and subspaces, we here give a brief overview of their basic definitions. 

A field $(\mathbb{K}, +_{\mathbb{K}}, \cdot_{\mathbb{K}})$ is a set equipped with two operations, field addition and multiplication, that satisfy the following: (i) both operations are associative and commutative, and multiplication is distributive over addition; (ii) both operations have identities, that is, there exist $0,1\in \mathbb K$ such that $1\cdot_{\mathbb{K}}a = a$ and $0+_{\mathbb{K}}a=a$ for any $a \in \mathbb{K}$; (iii) both operations have inverses: for any $a\in \mathbb{K}$, there are $b=(-a), c= a^{-1} \in \mathbb{K}$ such that $a+_{\mathbb{K}}b=0$ and $a\cdot_{\mathbb{K}}c = 1$. 
An example of a field is the set $\mathbb{Q}$ of rational numbers, with the usual operations.
A field is called \emph{finite} if it has a finite number of elements, $c = |\mathbb{K}|$. In this case, the field is often denoted as $\fc$, as it is known that all finite fields of a given cardinality are isomorphic to each other~\cite{Lidl:1997}. The set \zp of integers modulo a prime $p$ is a finite field. 

A vector space \bigspace over field $\mathbb{K}$ is a set equipped with two internal operations: vector addition and scalar multiplication, respectively $+:\bigspace\times \bigspace \rightarrow \bigspace$ and $\cdot:\mathbb K \times \bigspace \rightarrow \bigspace$. By definition, the operations satisfy the following properties:
\begin{enumerate}
    \item Vector addition is associative and commutative, and for this operation there exists the identity ($0\in \bigspace$ such that $v+0 = v$ for all $v\in \bigspace$) and the inverse (for any $v\in \bigspace$ there exists $w = -v \in \bigspace$ such that $v + w = 0$).
    \item Scalar multiplication is distributive with respect to vector addition, and has an identity element as well, which coincides with the identity of the field multiplication: $1\in \mathbb K$ is such that $1v=v$ for all $v\in \bigspace$. 
    \item Scalar multiplication is compatible with field multiplication: $(ab)v = a(bv)$ for any $a,b\in \mathbb K, v \in \bigspace$
    \item Scalar multiplication is distributive with respect to both vector addition and field addition. 
\end{enumerate} 
A vector space $\solution$ over the same field $\mathbb{K}$ is a vector \emph{subspace} of \bigspace if $\solution\subseteq \bigspace$.
A \emph{basis} of a vector space \bigspace is a set of vectors $b_1,...,b_n\in \bigspace$ such that any vector $v\in \bigspace$ can be expressed as a finite linear combination of the $b_is$ in a unique way: there exist $a_1,...,a_n\in \mathbb{K}$ such that $v= a_1b_1+...+a_nb_n$. Once a basis is fixed, we can uniquely represent the vector as its \emph{coordinates} given by the field coefficients with respect to the given basis: in the previous case, $v = (a_1,...,a_n)$.
All basis have the same cardinality, which is referred to as the \emph{dimension} of the a vector space. 

It is possible to  associate with each vector space \bigspace a unique  {\em canonical basis}. 
For instance, we can transform each basis $\cal B$  of \bigspace into the same unique canonical one by applying the Gauss–Jordan elimination algorithm to the matrix $M_{\cal B}$ whose rows are the vectors of $\cal B$.  This procedure transforms $M_{\cal B}$ into its {\em reduced-row-echelon form}~\cite{C81,L03}. 
Recall that a matrix is in reduced row echelon form if all of the following conditions are met:
\begin{enumerate} 
\item each leading coefficient (i.e., the first nonzero entry of each row) is equal to 1;
\item  rows of all zeros appear last;
\item for each pair of successive rows that are not all zeros, the leading coefficient of the
first row comes in an earlier column than the leading coefficient of the following
row;
\item  each pivotal column (i.e., a column that contains the leading coefficient of one row)
has only one nonzero entry.
\end{enumerate} 
Canonical bases of this type are easy to recognize: it is sufficient to check that each vector has a $1$ in the position of its own pivot and $0$ in the pivot positions of all the other vectors. This can be done by performing a number of vector operations proportional to the number of vectors in the basis, for example, by checking whether the arithmetic (non modular) sum of all vectors equals 1 in all pivot positions.

\section{A Motivational Application}
\label{section:motivation}

We recall here a problem related to the discovery of structural regularities in Boolean functions, which involves identifying a vector space of maximum dimension enclosed in a given set of vectors. This problem arises from the idea of extending a structural property known as autosymmetry from completely specified Boolean functions to incompletely specified ones.

A \emph{completely specified Boolean function} $f$ depending on $\ell$ input variables is a function $f:\{0,1\}^\ell \rightarrow \{0,1\}$ defined on all possible input configurations, while an \emph{incompletely specified Boolean function} $f$ in $\ell$ variables is a function $f:\{0,1\}^\ell\rightarrow\{0,1,-\}$, whose outputs can assume the values $0$, $1$, and the don't care value $-$. A don't care means that the value of the function, on that particular input configuration, could be either 0 or 1. 
The counter-images of 0, 1, and - are called {\em on-set}, {\em off-set}, and {\em dc-set}, respectively, and are denoted by $\fon$,   $\foff$,   and $\fdc$.  If $f$ is completely specified, then $\fdc = \emptyset$.

Autosymmetric functions are Boolean functions that exhibit special structural regularity easily expressed by exploiting the bitwise modulo 2 sum, and, in general, the algebraic structure of the domain $\{0,1\}^\ell$, viewed as a vector space over $\zdue$. They were introduced in~\cite{LP99} and further studied in~\cite{BCCM22, BCLP06, KS00a,KS00}, where it was shown how this regularity can be exploited to speed up the logic synthesis process and to derive compact logic representations, both for the classic and the quantum framework.   

Let $f:\{0,1\}^{\ell} \rightarrow \{0,1\}$ and  consider the set of vectors 
$$L_{f} = \{ \alpha \in \{0,1\}^\ell \, |\,  \forall\,w \in \fon,\  w \oplus \alpha \in  \fon \}.$$ 
We can notice that {\em (i)} the zero vector \mbox{\boldmath$0$} $\in L_{f}$;
  and {\em (ii)} if $\alpha_{1}, \alpha_{2}\in L_f$, then $\alpha_{1}\oplus \alpha_{2} \in L_f$. Thus, the set $L_{f}$  is a vector subspace of the domain $\{0,1\}^\ell$, called the \emph{vector space} of $f$.  Observe that $L_f$ contains $2^k$ vectors and it has dimension $k = \log_2 |L_f|$. 

We can now recall the definition of autosymmetry: 
\begin{definition}  
\label{def-k-auto} A completely specified Boolean function $f$ is \emph{$k$-autosymmetric}, $0\leq k \leq \ell$, if its vector space $L_{f}$ has dimension $k$.
A function $f$ is {\em autosymmetric} if it is $k$-autosymmetric with $k \ge 1$.
\end{definition}
Autosymmetric functions can be reduced to ``equivalent, but smaller'' functions. Intuitively, the reason is that $f$ takes a constant value on $L_f$, and it is also constant on any affine space over $L_f$. Since the whole space 
$\{0,1\}^\ell$ can be partitioned by the $2^{\ell-k}$ different, and  disjoint, affine spaces over $L_f$, the idea is to pick a single representative for each affine space, and reduce $f$ to a function $f_k$ over $\ell-k$ variables only
The function $f_k$ is called a {\em restriction} of $f$: $f_{k}$ is ``equivalent'' to, but smaller than $f$, and has $|\fon|/2^k$ points only, where $|\fon|$ denotes the cardinality of $\fon$.

Even if the set of autosymmetric functions is much smaller than the one containing all Boolean functions, a considerable amount of standard Boolean functions of practical interest falls in this set~\cite{BCCM22}. Thus, autosymmetry is a property that is frequent enough within Boolean functions to be worth studying.  
Moreover, autosymmetric functions can be efficiently identified among the set of completely specified Boolean functions, as their vector space and consequently the restriction $f_k$ can be derived in time polynomial in the dimension of some standard representations of the original function $f$~\cite{BCLP06}. 
This enables the exploitation of autosymmetry in the logic synthesis process: the idea is to minimize the restriction $f_k$, which depends on a smaller number of variables and has only $|\fon|/2^k$ points, and then to efficiently derive a minimal form for $f$ from a minimal form for $f_k$. 

Due to these properties---frequency, efficiency of the autosymmetry test, and the ability to derive compact logic representations in a shorter time---the notion of autosymmetry has been extended to incompletely specified functions in~\cite{BC21}. However, this extension has proven to be more demanding than expected.
Indeed, an incompletely defined function $f=(\fon, \fdc, \foff)$  with $t$ don't cares actually represents a family of $2^t$ different functions, where each function corresponds to one of the possible assignments of the values 0 and 1 to the $t$ don't cares. Thus, the idea is to select the function with the highest autosymmetry degree among the $2^t$ completely specified functions associated to $f$. 
In order to do so, the set
$$S_f = \{\alpha \in  \{0,1\}^\ell \ |\ \forall w \in \fon, \ w \oplus \alpha \in \fon \cup \fdc\}\,$$ is considered,
which is called   {\em closure set}  of $f$  and, unlike the case of completely specified functions, is not in general a vector space. 
Then, to maximize the autosymmetry degree of  the incompletely specified function $f$, it is necessary  to ``extract'' a vector space $L_f$ of maximum dimension contained in $S_f$. Observe that this is precisely an instance of the \textsc{Maximum Enclosed Space} problem, with ${\inputset} = S_f$,  ${\bigspace} = \{0,1\}^\ell$, and ${\solution}= L_f$.
In~\cite{BC21}, this task was solved using heuristic methods.

Finally, observe how this application requires the selective removal of elements from a given set of vectors in order to enhance the structural regularity of the domain.

\section{Additional Experimental Data}
\label{section:study_on_spaces}

\begin{figure}[h!]
    \centering
    \includegraphics[width=0.6\textwidth]{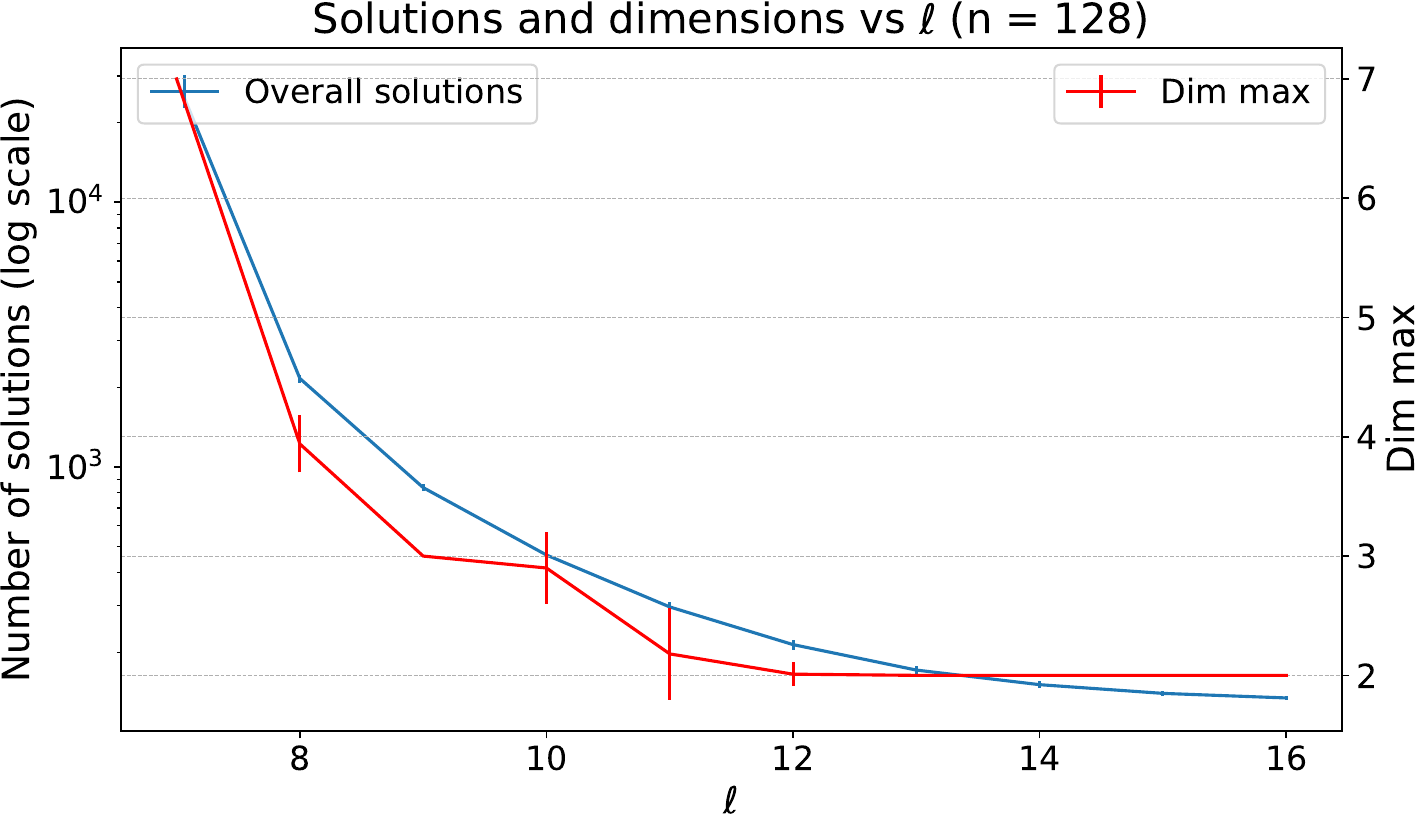}
    \caption{Number of enclosed spaces (log scale) and dimension of the maximum space for varying vector lengths $\ell$, with input size $n = 128$.}
    \label{fig:sol_vs_dim}
\end{figure}

The plot of Figure~\ref{fig:sol_vs_dim} shows that, for fixed cardinality of $\inputset$, as $\ell$ increases (and consequently the density of $\inputset$ within $\bigspace$ decreases), both the number of solutions and the size of the maximum solution decrease.

\end{document}